\documentclass[preprintnumbers,amsmath,amssymb,floatfix,10pt,twocolumn,prd,superscriptaddress,nofootinbib]{revtex4}
\usepackage{graphicx}
\usepackage{epsfig}
\usepackage{bm}
\usepackage{amsfonts}
\begin{document}
\title{Power-Law Entropy Corrected Holographic Dark Energy Model}
\author{Ahmad Sheykhi} \email{sheykhi@mail.uk.ac.ir}
\address{Department of Physics, Shahid Bahonar University, P.O. Box 76175, Kerman, Iran\\
         Research Institute for Astronomy and Astrophysics of Maragha (RIAAM), Maragha,
         Iran}
\author{Mubasher Jamil}
\email{mjamil@camp.nust.edu.pk} \affiliation{Center for Advanced
Mathematics and Physics, National University of Sciences and
Technology, H-12, Islamabad, Pakistan}

 \begin{abstract}
Among various scenarios to explain the acceleration of the universe
expansion, the holographic dark energy (HDE) model has got a lot of
enthusiasm recently. In the derivation of holographic energy
density, the area relation of the black hole entropy plays a crucial
role. Indeed, the power-law corrections to entropy appear in dealing
with the entanglement of quantum fields in and out the horizon.
Inspired by the power-law corrected entropy, we propose the
so-called ``power-law entropy-corrected holographic dark energy"
(PLECHDE) in this Letter. We investigate the cosmological
implications of this model and calculate some relevant cosmological
parameters and their evolution. We also briefly study the so-called
``power-law entropy-corrected agegraphic dark energy" (PLECADE).
\end{abstract}
\maketitle

\section{The Model\label{Int}}
One of the dramatic candidate for dark energy, that arose a lot of
enthusiasm recently, is the so-called ``holographic dark energy"
(HDE) proposal. This model is based on the holographic principle
which states that the number of degrees of freedom of a physical
system should scale with its bounding area rather than with its
volume \cite{Suss1} and it should be constrained by an infrared
cutoff \cite{Coh}. On this basis, Li \cite{Li} suggested the
following constraint on its energy density $\rho_D
\leq3c^2M^2_p/L^2$, the equality sign holding only when the
holographic bound is saturated. In this expression $c^2$ is a
dimensionless constant, $L$ denotes the IR cutoff radius and $M^2_p
=(8\pi G)^{-1}$ stands for the reduced Plank mass. Based on
cosmological state of holographic principle, proposed by Fischler
and Susskind \cite{Suss2}, the HDE models have been proposed and
studied widely in the literature (see e.g.
\cite{Huang0,Hsu,wang1,HDE,sheykhi} and references therein). The HDE
model has also been tested and constrained by various astronomical
observations \cite{Xin,Feng} as well as by the anthropic principle
\cite{Huang1}. It is fair to claim that simplicity and reasonability
of HDE model provides more reliable frame to investigate the problem
of dark energy rather than other models proposed in the literature.
For example, the coincidence problem can be easily solved in some
models of HDE based on the fundamental assumption that matter and
HDE do not conserve separately \cite{pav1}.

It is worthy to note that the definition and derivation of
holographic energy density ($\rho_{D } =3c^2M^2_p/L^2$) depends on
the entropy-area relationship $S\sim
 A \sim L^2$ of black holes, where $A$ represents
the area of the horizon \cite{Coh}. However, this definition of HDE
can be modified due to the power-law corrections to entropy which
appear in dealing with the entanglement of quantum fields in and out
the horizon \cite{Sau}. The power-law corrected entropy takes the
form \cite{pavon1}
\begin{equation}\label{S}
S=\frac{A }{4G}\left[1-K_{\alpha}A^{1-\alpha/2}\right],
\end{equation}
where $\alpha$ is a dimensionless constant whose value is
currently under debate, and
\begin{equation}
\label{K} K_{\alpha}=\frac{\alpha
(4\pi)^{\alpha/2-1}}{(4-\alpha)r_c^{2-\alpha}},
\end{equation}
where $r_c$ is the crossover scale. The second term in Eq. (1) can
be regarded as a power-law correction to the area law, resulting
from entanglement, when the wave-function of the field is chosen to
be a superposition of ground state and exited state \cite{Sau}.
{The entanglement entropy of the ground state obeys
the Hawking area law. Only the excited state contributes to the
correction, and more excitations produce more deviation from the
area law \cite{sau1} (also see \cite{sau2} for a review on the
origin of black hole entropy through entanglement).} This lends
further credence to entanglement as a possible source of black hole
entropy. The correction term is also more significant for higher
excitations \cite{Sau}. It is important to note that the correction
term falls off rapidly with $A$  (see the discussion in favor of
$\alpha>2$ in the below) and hence in the semi-classical limit
(large $A$) the area law is recovered. {So for large
black holes the correction term falls off rapidly and the area law
is recovered, whereas for the small black holes the correction is
significant. This can be interpreted as follows: for large area,
i.e., at low energies, it is difficult to excite the modes and
hence, the ground state modes contribute to most of the entanglement
entropy. However, for small horizon area, a large number of field
modes can be excited and contribute significantly to the correction
causing large deviation from the area law.}

Inspired by the power-law corrected entropy  relation (\ref{S}), and
following the derivation of HDE \cite{Gub} and entropy-corrected
holographic dark energy (ECHDE) \cite{Wei} , we can easily obtain
the energy density of the so-called ``power-law entropy-corrected
holographic dark energy" (PLECHDE), namely
\begin{equation}\label{rhoD}
\rho _{D }=3c^2M_{p}^{2}L^{-2}-\beta M_{p}^{2}L^{-\alpha}.
\end{equation}
In the special case  $\beta=0$, the above equation yields the
well-known holographic energy density. The significant of the
corrected term in various regions depends on the value of
$\alpha$. When $\alpha=2$ the two terms can be combined and one
recovers the ordinary HDE density. Let us consider the case with
$\alpha>2$ and $\alpha<2$ separately.  In the first case where
$\alpha>2$ the corrected term can be comparable to the first term
only when $L$ is very small. Indeed, it was argued that $\alpha$
should be ranges as $2<\alpha<4$ \cite{Sau}. However, the
satisfaction of the generalized second law of thermodynamics for
the universe with the power-law corrected entropy (1) implies that
the case $\alpha<2$   should be rejected \cite{pavon1}.

\section{Non-interacting case\label{HDE}}
We consider the non-flat Friedmann-Robertson-Walker (FRW) universe
which is described by the line element
\begin{eqnarray}
 ds^2=dt^2-a^2(t)\left(\frac{dr^2}{1-kr^2}+r^2d\Omega^2\right),\label{metric}
 \end{eqnarray}
where $a(t)$ is the scale factor, and $k$ is the curvature
parameter with $k = -1, 0, 1$ corresponding to open, flat, and
closed universes, respectively. The corresponding Friedmann
equation takes the form
\begin{eqnarray}\label{Fried}
H^2+\frac{k}{a^2}=\frac{1}{3M_{p}^{2}} \left( \rho_m+\rho_D
\right),
\end{eqnarray}
where $\rho_m$ and $\rho_D$ are the energy density of dark matter
and dark energy, respectively.

It is important to note that in the literature, various scenarios of
HDE have been studied via considering different system's IR cutoff.
In the absence of interaction between dark matter and dark energy in
flat universe, Li \cite{Li} discussed three choices for the length
scale $L$ which is supposed to provide an IR cutoff. The first
choice is the Hubble radius, $L=H^{-1}$ \cite{Hsu}, which leads to a
wrong equation of state, namely that for dust. The second option is
the particle horizon radius. In this case it is impossible to obtain
an accelerated expansion. Only the third choice, the identification
of $L$ with the radius of the future event horizon gives the desired
result, namely a sufficiently negative equation of state to obtain
an accelerated universe. However, as soon as an interaction between
dark energy and dark matter is taken into account, the first choice,
$L=H^{-1}$, in flat universe, can simultaneously drive accelerated
expansion and solve the coincidence problem \cite{pav1}. It was also
argued \cite{sheykhi} that in a non-flat universe the natural choice
for IR cutoff could be the apparent horizon radius
$\tilde{r}_A={1}/{\sqrt{H^2+k/a^2}}$ provided the interaction is
taken into account. In recent years, some new infrared cut-offs have
also been proposed in the literature. In \cite{GO}, the authors have
added the square of the Hubble parameter and its time derivative
within the definition of holographic dark energy. While in
\cite{Sad}, the authors proposed a linear combination of particle
horizon and the future event horizon. In this section, following
\cite{wang2}, as system's IR cutoff we choose the radius of the
event horizon measured on the sphere of the horizon, defined as
\begin{equation}\label{L}
L=ar(t),
\end{equation}
where the function $r(t)$ can be obtained from the following
relation
\begin{equation}
 \int_{0}^{r(t)}{\frac{dr}{\sqrt{1-kr^2}}}=\int_{0}^{\infty}{\frac{dt}{a}}=\frac{R_h}{a}.
\end{equation}
Solving the above equation for the general case of the non-flat
FRW universe, we have
\begin{equation}
r(t)=\frac{1}{\sqrt{k}}\sin y,\label{rt}
\end{equation}
where $y=\sqrt{k} R_h/a$. We also define as usual, the fractional
energy densities such as
\begin{eqnarray}
\Omega_m&=&\frac{\rho_m}{\rho_{\mathrm{cr}}}=\frac{\rho_m}{3M_{p}^2
H^2}, \label{Omegam} \\
\Omega_k&=&\frac{\rho_k}{\rho_{\mathrm{cr}}}=\frac{k}{H^2 a^2},\label{Omegak} \\
\Omega_D&=&\frac{\rho_D}{\rho_{\mathrm{cr}}}=\frac{\rho_D}{3M_{p}^2
H^2}. \label{OmegaD}
\end{eqnarray}
Now we can rewrite the  Friedmann equation in the following form
\begin{eqnarray}\label{Fried2}
1+\Omega_k=\Omega_m+\Omega_D.
\end{eqnarray}
Using the definitions of $\Omega_D$ and $\rho_D$, we obtain a
useful relation
\begin{eqnarray}
HL=\sqrt{\frac{3c^2-\beta L^{2-\alpha}}{{3\Omega_D}}}. \label{HL}
\end{eqnarray}
Taking derivative with respect to the cosmic time $t$ from Eq.
(\ref{L}) and using Eqs. (\ref{rt}) and (\ref{HL}) we obtain
\begin{eqnarray}
\dot{L}=HL+a\dot{r}(t)=\sqrt{\frac{3c^2-\beta
L^{2-\alpha}}{{3\Omega_D}}}-\cos y. \label{Ldot}
\end{eqnarray}
Consider the FRW universe filled with dark energy and pressureless
matter which evolves according to their conservation laws
\begin{eqnarray}
&&\dot{\rho}_D+3H\rho_D(1+w_D)=0,\label{consq}\\
&&\dot{\rho}_m+3H\rho_m=0, \label{consm}
\end{eqnarray}
where $w_D$ is the equation of state parameter of dark energy.
Differentiating (\ref{rhoD}) with respect to time and using Eq.
(\ref{Ldot}) we find
\begin{eqnarray}
\dot{\rho}_D&=&\left(-6c^2M_{p}^2L^{-3}+\alpha \beta
M_{p}^2L^{-\alpha-1}\right)\nonumber\\&&\times\left[\sqrt{\frac{3c^2-\beta
L^{2-\alpha}}{{3\Omega_D}}}-\cos y\right]\label{rhodot}.
\end{eqnarray}
Inserting this equation in conservation law (\ref{consq}), we
obtain the equation of state parameter
\begin{eqnarray}
w_D&=&-1+\frac{1}{3}\left[\frac{6c^2-\alpha \beta
L^{2-\alpha}}{3c^2- \beta
L^{2-\alpha}}\right]\nonumber\\&&\times\left[1-
\sqrt{\frac{{3\Omega_D}}{3c^2-\beta L^{2-\alpha}}}\cos
y\right]\label{wD}.
\end{eqnarray}
It is important to note that in the limiting case  $\beta=0$ Eq.
(\ref{wD}) reduces to its respective expression in ordinary HDE
\cite{Huang}
\begin{eqnarray}
w_D=-\frac{1}{3}-\frac{2\sqrt{\Omega_D}}{3c}\cos y\label{wDstand}.
\end{eqnarray}
For completeness, we give the deceleration parameter
\begin{eqnarray}
q=-\frac{\ddot{a}}{aH^2}=-1-\frac{\dot{H}}{H^2},
\end{eqnarray}
which combined with the Hubble parameter and the dimensionless
density parameters form a set of useful parameters for the
description of the astrophysical observations. Taking the time
derivative of the Friedmann equation (\ref{Fried}) and using Eqs.
(\ref{Fried2}), (\ref{consq}) and (\ref{consm}) we obtain
\begin{eqnarray}
q=\frac{1}{2}\left[1+\Omega_k+3\Omega_D w_D\right]\label{q1}.
\end{eqnarray}
Substituting $w_D$ from Eq. (\ref{wD}), we get
\begin{eqnarray}
q&=&\frac{1}{2}\Big[1+\Omega_k-3\Omega_D+\Omega_D
\Big(\frac{6c^2-\alpha \beta L^{2-\alpha}}{3c^2- \beta
L^{2-\alpha}}\Big)\nonumber\\&&\times\Big(1-
\sqrt{\frac{{3\Omega_D}}{3c^2-\beta L^{2-\alpha}}}\cos y\Big)
\Big]\label{q4}.
\end{eqnarray}
When $\beta=0$, Eq. (\ref{q4}) restores the deceleration parameter
for standard HDE model \cite{wang2}
\begin{eqnarray}
q=\frac{1}{2}(1+\Omega_k)-\frac{\Omega_D}{2}-\frac{\Omega^{3/2}_D}{c}\cos
y\label{q3}.
\end{eqnarray}
%%%%%%%%%%%%%%%%%%%%%%%%%%%%%%%%%%%%%%%%%%%%%%%%%%%%%%%%%
\section{Interacting case\label{INTHDE}}

The above study can also be performed for the interacting case. In
the absence of a symmetry that forbids the interaction there is
nothing, in principle, against it. Indeed, microphysics seems to
allow enough room for the interacting \cite{Dor}. Taking the
interaction into account, the continuity equations read
\begin{eqnarray}
&&\dot{\rho}_D+3H\rho_D(1+w_D)=-Q,\label{consq2},\\
&&\dot{\rho}_{m}+3H\rho_{m}=Q, \label{consm2},
\end{eqnarray}
where $Q$ represents the interaction term which can be, in general,
an arbitrary function of cosmological parameters like the Hubble
parameter and energy densities $Q (H\rho_{m},H\rho_{D})$. The
simplest choice is $Q =3b^2 H(\rho_{m}+\rho_{D})$ with $b^2$ is a
coupling constant \cite{Ame,Zim}, although more general
phenomenological interaction terms can be used \cite{jamil}. The
positive $b^2$ is responsible for the transition from dark energy to
matter and vice versa for negative $b^2$. Sometimes this constant is
taken in the range $[0, 1]$  \cite{zhang}. Note that if $b^2=0$ then
it represents the non-interacting FRW model while $b^2=1$ yields
complete transfer of energy from dark energy to matter. Recently, it
is reported that this interaction is observed in the Abell cluster
A586 showing a transition of dark energy into dark matter and vice
versa \cite{berto1}. Observations of cosmic microwave background and
galactic clusters show that the coupling parameter $b^2< 0.025$,
i.e. a small but positive constant of order unity \cite{ich}, a
negative coupling parameter is avoided due to violation of
thermodynamical laws. {Therefore the theoretical interacting models
are phenomenologically consistent with the observations.} It should
be noted that the ideal interaction term must be motivated from the
theory of quantum gravity. In the absence of such a theory, we rely
on pure dimensional basis for choosing an interaction $Q$.

Inserting Eq. (\ref{rhodot}) in  (\ref{consq2}) and using relation
(\ref{HL})  we obtain the equation of state parameter
\begin{eqnarray}
w_D&=&-1+\frac{1}{3}\left(\frac{6c^2-\alpha \beta
L^{2-\alpha}}{3c^2- \beta L^{2-\alpha}}\right)\left[1-
\sqrt{\frac{{3\Omega_D}}{3c^2-\beta L^{2-\alpha}}}\cos
y\right]\nonumber\\&&-
\frac{b^2(1+\Omega_k)}{\Omega_D}\label{wDInt}.
\end{eqnarray}
If we define, following \cite{Setare1}, the effective equation of
state parameter as
\begin{eqnarray}\label{wef}
w^{\mathrm{eff}}_D=w_D+\frac{\Gamma}{3H},
\end{eqnarray}
Here $\Gamma=3b^2(1+u)H$, where $u= \rho_m/\rho_D$ is the energy
density ratio of two dark components. Then, the continuity equation
(\ref{consq2}) for the dark energy can be written in the standard
form
\begin{eqnarray}
&&\dot{\rho}_D+3H\rho_D(1+w^{\mathrm{eff}}_D)=0.\label{consqeff}
\end{eqnarray}
Substituting Eq. (\ref{wDInt}) in  Eq. (\ref{wef}), we find
\begin{eqnarray}\label{wDeff}
w^{\mathrm{eff}}_D&=&-1+\frac{1}{3}\left(\frac{6c^2-\alpha \beta
L^{2-\alpha}}{3c^2- \beta
L^{2-\alpha}}\right)\nonumber\\&&\times\left[1-
\sqrt{\frac{{3\Omega_D}}{3c^2-\beta L^{2-\alpha}}}\cos y\right]
\end{eqnarray}
Finally, we examine the deceleration parameter. Substituting $w_D$
from Eq. (\ref{wDInt}) in Eq. (\ref{q1}), we get
\begin{eqnarray}
q&=&\frac{1}{2}\Big[1+\Omega_k-3\Omega_D+\Omega_D
\Big(\frac{6c^2-\alpha \beta L^{2-\alpha}}{3c^2- \beta
L^{2-\alpha}}\Big)\nonumber\\&&\times\Big(1-
\sqrt{\frac{{3\Omega_D}}{3c^2-\beta L^{2-\alpha}}}\cos y\Big)
-3b^2(1+\Omega_k)\Big]\label{q2}.
\end{eqnarray}

\section{PLECHDE with HUBBLE HORIZON AS IR CUT-OFF\label{HDEH1}}

In this section we consider PLECHDE with $L=H^{-1}$ as an IR-cutoff
in a flat FRW universe. This cutoff is particularly
relevant for the very early universe undergoing a hypothetical phase
of inflation - a very brief period of exponential expansion. After
the end of inflationary phase, the universe evolved subsequently
through the radiation and matter phases. In the last two stages, the
Hubble horizon is replaced with the future event horizon $R_h$ as a
dynamical cutoff. Consequently the power-law-correction to HDE
becomes negligible until the beginning of late time cosmic
acceleration. Note that the HDE with the Hubble Horizon as cutoff
can not generate late time accelerated expansion and only dynamical
future event horizon can serve this purpose. Therefore the choice of
Hubble horizon to generate cosmic acceleration is restricted to the
early universe. In this case, the energy density of PLECHDE can be
rewritten as
\begin{equation}\label{rhoH}
\rho _{D }=3c^2M_{p}^{2}H^{2}-\beta M_{p}^{2}H^{\alpha}.
\end{equation}
Differentiating with respect to time we obtain
\begin{equation}\label{rhodotH}
\dot{\rho} _{D }=\dot{H}H M_{p}^{2}\left(6c^2-\alpha\beta
H^{\alpha-2}\right).
\end{equation}
Taking the time derivative of  Friedmann equation (\ref{Fried})
for the flat universe ($k=0$) and using the continuity equations
(\ref{consq}) and (\ref{consm}), we get
\begin{equation}\label{dotH}
\dot{H}=-\frac{\rho_D}{2M_{p}^{2}}  \left(1+u+w_D\right).
\end{equation}
Inserting Eqs. (\ref{rhodotH}) and (\ref{dotH}) in Eq. (\ref{consq})
we can easily obtain the equation of state parameter of PLECHDE
\begin{equation}\label{wDH}
w_D=-1-\frac{(6c^2-\alpha \beta H^{\alpha-2})u}{6(c^2-1)-\alpha
\beta H^{\alpha-2}}.
\end{equation}
It is worth noting that in the absence of correction term
($\beta=0$) the above equation reduces to
\begin{equation}\label{wDH2}
w_D=-1-\frac{c^2}{c^2-1}u,
\end{equation}
while from the Friedmann equation we find $u=1/c^2-1$.
Substituting this relation in Eq. (\ref{wDH2}) we obtain $w_D=0$,
which is a wrong equation of state for dark energy and cannot
derive the acceleration of the universe expansion \cite{Hsu}.
However, as one can see from Eq. (\ref{wDH}) in the presence  of
the power-law correction term, the identification of IR-cutoff
with Hubble radius, $L=H^{-1}$, can lead to accelerated expansion.

\section{power-law entropy-corrected new agegraphic dark
energy\label{HDEH}}

A so-called agegraphic dark energy (ADE) is originated from
uncertainty relation of quantum mechanics together with the
gravitational effect in general relativity. The ADE model assumes
that the observed DE comes from the spacetime and matter field
fluctuations in the universe.  However, the original ADE \cite{Cai1}
model had some difficulties. For example it suffers from the
difficulty to describe the matter-dominated epoch, there is no
inflation attractor using ADE even if the entropy corrections are
applied to it and its density falls as the universe expands unlike
typical dark energy candidates including cosmological constant and
quintessence \cite {Wei}. Therefore, a new model of ADE was proposed
by Wei and Cai \cite{Wei2}, while the time scale was chosen to be
the conformal time instead of the age of the universe. The new ADE
(NADE) contains some new features different from the original ADE
and overcome some unsatisfactory points. The ADE models have been
examined and studied in ample detail (see e.g.
\cite{age,shey1,shey2,karami2} and Refs. therein). The energy
density of the NADE is given by \cite{Wei2}
\begin{equation}
\rho_{D} = \frac{3n^2{M_P^2}}{\eta^2},\label{ecnade}
\end{equation}
where the conformal time $\eta$ is given by
\begin{equation}
\eta=\int_0^{a}\frac{{\rm d}a}{Ha^2}.\label{eta}
\end{equation}
Here, we would like to propose the so-called ``power-law
entropy-corrected agegraphic dark energy" (PLECADE) whose $L$ in
Eq. (\ref{rhoD}) is chosen to be the conformal time $\eta$.
Therefore, we write down the energy density of PLECADE as
\begin{equation}\label{rhoeta}
\rho _{D }=3n^2M_{p}^{2}{\eta}^{-2}-{\beta}
M_{p}^{2}{\eta}^{-\alpha}.
\end{equation}
To be more general we consider the interacting case. Taking the
time  derivative of Eq. (\ref{rhoeta}) and using the relation
$\dot{\eta}=1/a$ we find
\begin{equation}\label{rhoetad}
\dot{\rho} _{D
}=-\frac{6n^2M_{p}^{2}}{a{\eta}^{3}}+\frac{\alpha\beta
M_{p}^{2}}{a{\eta}^{\alpha+1}}.
\end{equation}
Using definition (\ref{OmegaD}) as well as (\ref{rhoeta}) we
obtain
\begin{eqnarray}
H\eta=\sqrt{\frac{3n^2-\beta \eta^{2-\alpha}}{{3\Omega_D}}}.
\label{Heta}
\end{eqnarray}
Substituting Eq. (\ref{rhoetad}) in (\ref{consq2}) and using
relation (\ref{Fried2}) and (\ref{Heta}) we find the equation of
state parameter of PLECADE as
\begin{eqnarray}
w_D&=&-1+\frac{\sqrt {3\Omega_D}}{3a\sqrt{3n^2-\beta
\eta^{2-\alpha}}}\left(\frac{6n^2-\alpha \beta\eta^{2-\alpha}}{3n^2-
\beta \eta^{2-\alpha}}\right)\nonumber\\&&-
\frac{b^2(1+\Omega_k)}{\Omega_D}\label{wDIntADE}.
\end{eqnarray}
In the limiting case $\beta=0$, one recovers the equation of state
parameter of usual NADE, namely

\begin{eqnarray}
w_D=-1+\frac{2\sqrt {\Omega_D}}{3na}-
\frac{b^2(1+\Omega_k)}{\Omega_D}\label{wDIntADE}.
\end{eqnarray}

%%%%%%%%%%%%%%%%%%%%%%%%%%%%%%%%%%%%%%%%%%%%%%%%%%%%%%%%%%%%%%%%%%
\section{CONCLUSION\label{CONC}}
It has been shown that the origin of black hole entropy may lie in
the entanglement of quantum fields between inside and outside of
the horizon \cite{Sau}. Since the modes of gravitational
fluctuations in a black hole background behave as scalar fields,
one is able to compute the entanglement entropy of such a field,
by tracing over its degrees of freedom inside a sphere. In this
way the authors of \cite{Sau} showed that the black hole entropy
is proportional to the area of the sphere when the field is in its
ground state, but a correction term proportional to a fractional
power of area results when the field is in a superposition of
ground and excited states. For large horizon areas, these
corrections are relatively small and the area law is recovered.

Motivated by the power-law corrected entropy, we  proposed the
so-called ``power-law entropy-corrected holographic dark energy"
(PLECHDE) in this Letter. We calculated some relevant cosmological
parameters such as the equation of state and deceleration parameter
of this model. We also extended our study by incorporating the
interaction term in the this model and obtained the equation of
state for the interacting power-law entropy corrected holographic
dark energy density in a non-flat universe. Interestingly enough, we
found that in the presence of the power-law correction term, the
identification of IR-cutoff with Hubble radius, $L=H^{-1}$, can
drive the accelerated expansion. This is in contrast to the ordinary
HDE where $w_D=0$ if one choose $L=H^{-1}$ in the absence of
interaction \cite{Hsu}. Finally, we performed the study for the
power-law entropy corrected new agegraphic dark energy and obtained
its equation of state parameter. Here the phantom crossing scenario
becomes plausible if $n>0$ and scale factor is large (or small
redshift $z\simeq1$). We plan to extend this work by developing
correspondences between the PLECHDE and various other dark energy
candidates modeled by using scalar fields. Besides the Einstein's
gravity, it can be extended to Brans-Dicke chameleon cosmology and
$f(R)$ gravity. Moreover it would be interesting to differentiate
PLECHDE from other dark energy candidates by checking the
corresponding statefinder parameters. These issues are now under
consideration and will be addressed elsewhere.

%%%%%%%%%%%%%%%%%%%%%%%%%%%%%%%%%%%%%%%%%%%%%%%%%%%%%%%%%%%%%%%%%%%%%%%
\acknowledgments{We are grateful to Professors H. Wei and B. Wang
for helpful discussions. This work has been supported by Research
Institute for Astronomy and Astrophysics of Maragha, Iran.}

\end{document}